# A Novel Approach in Strategic Planning of Power Networks Against Physical Attacks

Hamzeh Davarikia [1*], Masoud Barati [2], Mustafa Al-Assad [3], Yupo Chan [3]

[1] McNeese State University, LA, USA
[2] University of Pittsburgh, PA, USA
[3] University of Arkansas at Little Rock, AR, USA

**Abstract**

The reported work points at developing a practical approach for power transmission planners to secure power networks from potential deliberate attacks. We study the interaction between a system planner (defender) and a rational attacker who threatens the operation of power grid. In addition to the commonly used hardening strategy for protecting the network, a new sort of resources is introduced under the deception concept. Feint and deception are acknowledged as effective tools for misleading the attacker in strategic planning. To this end, the defender deception is mathematically formulated by releasing misinformation about his plan in the shared cognition-based model. To reduce the risk of damage in case of deception failure, preemptive-goal programming is utilized to prioritize the hardening strategy for the vital components. Furthermore, the "value of posturing" is introduced which is the benefits that the deception brings to the system. The problems are formulated as tri-level mixed-integer linear programming and solved by constraint-and-column generation method. Comprehensive simulation studies performed on WSCC 9-bus and IEEE 118-bus systems indicate how the defender will save significant cost from protecting his network with posturing rather than hardening and the proposed approach is a promising development to ensure the secure operation of power networks.

**Nomenclature**

**Indices and sets:**

| | |
|---|---|
| $i$ | Bus index; $i \in B$. |
| $g$ | Generator index; $g \in G$. |
| $ij$ | Transmission line index; $(i,j) \in L$. |
| $B$ | Set of indices of buses. |
| $G$ | Set of indices of generators. |
| $L$ | Set of indices of transmission lines. |
| $Gb_i$ | Set of indices of generators connected to bus $i$. |

**Continues Variables:**

| | |
|---|---|
| $P_g^G$ | The power output of generator $g$. |
| $P_i^{Sh}$ | Load shedding at bus $i$. |
| $\theta_i$ | Voltage angle on bus $i$. |
| $F_{ij}$ | Power flow through line $(i,j)$. |

**Integer variables:**

| | |
|---|---|
| $w_g^D$ | Units of capacity $q_g$ added to generator $g$. |
| $v_{ij}^D$ | Units of capacity $c_{ij}$ added to the transmission line $(i,j)$. |

**Binary variables:**

| | |
|---|---|
| $x_i^D$ | 1 if bus $i$ is protected and 0 otherwise. |
| $z_g^D$ | 1 if generator $g$ is protected and 0 otherwise. |
| $y_{ij}^D$ | 1 if transmission line $(i,j)$ is protected and 0 otherwise. |
| $x_i^f$ | 1 if the defender generates false data on protection status $x_i^D$ and 0 otherwise. |
| $z_g^f$ | 1 if the defender generates false data on protection status $z_g^D$ and 0 otherwise. |
| $y_{ij}^f$ | 1 if the defender generates false data on protection status $y_{ij}^D$ and 0 otherwise. |
| $x_i^{Df}$ | The protection status of $x_i^D$ observed by the attacker. |
| $z_g^{Df}$ | The protection status of $z_g^D$ observed by the attacker. |
| $y_{ij}^{Df}$ | The protection status of $y_{ij}^D$ observed by the attacker. |
| $x_i^A$ | 1 if bus $i$ is attacked and 0 otherwise. |
| $z_g^A$ | 1 if generator $g$ is attacked and 0 otherwise. |
| $y_{ij}^A$ | 1 if transmission line $(i,j)$ is attacked and 0 otherwise. |

**Parameters:**

| | |
|---|---|
| $P_i^D$ | Summation of loads connected to bus $i$. |
| $C_i^{Sh}$ | Load-shedding cost at bus $i$. |
| $B_{ij}$ | Susceptance for transmission line $(i,j)$. |
| $\overline{F_{ij}}$ | Maximum capacity of transmission line $(i,j)$. |
| $c_{ij}$ | Size of capacity increments that can be added to transmission line $(i,j)$. |
| $C_g$ | The production cost of generating unit $g$. |
| $q_g$ | Size of unit increments that can be added to generator $g$. |
| $\overline{P_g^G}$ | Maximum generation capacity of generator $g$. |
| $R^{PB}$ | Number of defender's resource for protecting buses. |
| $R^{PG}$ | Number of defender's resource for protecting generators. |
| $R^{PL}$ | Number of defender's resource for protecting transmission lines. |
| $R^{RL}$ | Defender's resource for the maximum capacity of transmission line reinforcement. |
| $R^{RG}$ | Defender's resource in terms of the maximum capacity of generators reinforcement. |
| $R^{FB}$ | Number of defender's resource for generating false data on protection status of buses |
| $R^{FG}$ | Number of defender's resource for generating false data on protection status of generators. |
| $R^{FL}$ | Number of defender's resource for generating false data on protection status of lines. |
| $R^{AB}$ | Number of attacker's resource for simultaneously attacking to buses. |
| $R^{AG}$ | Number of attacker's resource for simultaneously attacking to generators. |
| $R^{AL}$ | Number of attacker's resource for simultaneously attacking to transmission lines. |

## 1. Introduction

The electric power grid is recognized as one of the most sophisticated systems made by humanity, and its importance is acknowledged when there is a threat or a costly power outage. As a recent example, a crafted incident involved a sophisticated sniper attack on a substation in Metcalf, California in April 2013. Due to this attack, the substation was down for almost a month, and the cost of the damage was estimated to be $15.4 million [1]. According to the U.S. Department of Homeland Security, from 2011 to 2014, the national power grid came under physical or cyber attack once every four days [2]. Due to this growing cyber-physical threat against power network, beefing up the power grid resilience is an essence. This paper provides a novel defensive scheme to protect the most vital elements in the power system against malicious attacks. To this end, the defender's deception under the shared cognition concept [3] is introduced, and the monetary worth of the misleading the attacker is calculated.

Accepted to be published in Journal of Electric Power Systems Research, 2018The conventional approach to address the planning against deliberate attack is to establish a sequential game among defender and the attacker. This can be formulated as the bi-level attacker-defender (AD) or tri-level defender-attacker-defender (DAD) Stackelberg game. Reference [4] proposed a comprehensive AD approach to identify critical system components, and the solution obtained by a heuristic algorithm where the optimality is uncertain. In [5], the same authors proposed the "Global Benders Decomposition" algorithm to replace the heuristic approach. However, discontinuity of sub-problem may lead to an inaccurate solution. As a more classical solution approach references [6, 7] applied, respectively, the strong duality theorem and first-order optimality conditions known as Karush-Kuhn-Tucker (KKT) conditions to transform the bi-level AD model into a one-level problem and implemented the mixed-integer linear programming (MILP) solvers.

Yao et al. [8] was the pioneer of applying tri-level modeling approach to develop the DAD model in power system, where the solution achieved by a decomposition-based method. Romero et al. [9], introduced a comprehensive tri-level mixed-integer non-linear model and utilized the heuristic approach as the solution strategy. In the modest model, Wu et al. [10], proposed a tri-level model and applied an instance of Bender's decomposition technique using primal cuts to achieve a more effective result. Xiang et al. [11] proposed the tri-level DAD model and the solution obtained by the robust optimization approach. The attacker's damage is quantified by the load curtailment in [12], where the risk optimization method utilized to minimize the disruption's consequence. Mitigating the impact of cyber-physical attack discussed in [13], and the authors examine the consequence of interrupting the transmission lines or generators in addition to load redistribution attacks.

Hardening refers to the protection status of an element that makes the component invulnerable to damage and considered as the commonly used approach for protecting the critical assets in resilience improvement and interdiction problems in power networks. Lin et al. [14] proposed attack-resilient planning for the distribution networks through a tri-level DAD interdiction problem which allocates the hardening resources among the system components where the reconfiguration and DG islanding are considered. The hardening concept is adopted to equip the distribution networks against the natural hazards in [15, 16], where the hardened components help the distribution system to maintain the minimum load shed when tolerating the extreme events. The hardening measures are utilized to facilitate the restoration process in distribution networks after the damage occurred [17, 18]. This achieved by protecting the key equipment in the grid when the system planner has limited resources for hardening. Improving the bulk power system resilience against physical attack is pursed in [19, 20], where the authors formulated the optimization problem to locate the hardening strategies for the most vulnerable component.

Going beyond the conventional approach in the literature, the shared cognition-based model implies soft factors in the mathematical formulation, which explains the role of information in the team and organizational performance [21]. One aspect of shared cognition is the defender's deception, which is the actions executed to mislead the adversary deliberately. It creates uncertainty and confusion against the adversary's efforts to found situational awareness and to effect and misdirect adversary perceptions and decision processes. Aside from outright fortification, disseminating false information for protecting the network is a positive strategy [22], where the feint and deception have become the commonly used tactic in conflicts [23]. For example, the company can consider when he is in the lack of hardening resources, whereby at least it can deceive the attacker into attacking less essential facilities. However, how many facilities need hardening, and how much false information is required to generate? Another common example is the police car, where you can find some parked police car without any cops in them. They are used for posturing and improving security. But how police station should allocate their limited cars to maximize the security?

Deliberate deception indicates strategic efforts strived at misleading opponents from the actual strategic plan of the defender. The degree and type of deception may differ from the simple masking of trivial data to outright lies and propagating disinformation. Crawford et al. [24] emphasized the importance of lying for strategic benefit about planned tactics and proposed a game-theoretic approach to misdirect the enemies. Zhuang et al. [25] developed a model to explore the effectiveness of the deception strategy in disclosing the planned decisions. The same authors developed the signaling game under the information secrecy and deception in [26], where the defender could keep his strategy secret or mislead the attacker. This type of problem can also be viewed as the case that the attacker is not strategic or rational enough [27]. The advantage of secrecy and feint from the Homeland-Security point of view is discussed in [28], where the authors conclude that the deception and secrecy about data disclosure is an advantageous strategy for a planner to increase the effectiveness of defensive tactics. Following the same concept of the literature mentioned above, Ma et al. [29] adopted the deception as a defensive strategy and assumed that the defender could mislead the attacker into using incorrect cost functions that quantify the damage of load shedding. In spite of a wealth of research on utilizing deception and secrecy on homeland security and critical infrastructure security enhancements, but it is hard to find the example of such strategies since these tactics are often classified [28]. While the reviewed works developed some game-theoretic approaches to analyzing the deception effectiveness on the planned strategies, their method neither promises a resource allocation-based approach to distribute the limited deception resource among the component nor determines the facilities deserved for the posturing strategy.

In this paper, a new approach for deception modeling is introduced, and the proposed method is appropriated in a resource allocation-based DAD model to find the best posturing strategies on the protection status of the equipment. To avoid the risk of deception failure, a preemptive goal programming method is employed to prioritize the hardening resources to the vital components. It is also essential to estimate the value of the hardening as well as that associated with deception strategy; representing new findings not reported in the literature. The cost savings provide the valuation of our model. By experimenting with the number of hardened facilities and deceptions, the defender can see how many of such resources are required to deviate the attacker from the most critical assets. Calculating the amount of savings in dollars is of great importance to governments and companies who can now adopt judicious, cost-effective steps to prevent malicious attacks based on their valuation.

The prominent contributions and novelty of the present work are outlined in the ensuing parts: 1) providing the mathematical programming approach for posturing and deception in terms of shared cognition paradigm; 2) developing a resource allocation-based DAD model to find the best feint strategy; 3) quantifying the value of deception in a game-theoretic context; 4) employing the preemptive goal programming approach to distinguish between hardening/reinforcement and posturing strategies to adopt the effective defense for the vital components. The rest of this paper is organized as follow: problem formulation is discussed in Section 2 followed by the solution strategy in Section 3. Extensive simulations elucidate the efficacy of the proposed technique on different test systems in Section 4. The paper is concluded with discussions of opportunities for extensions of the proposed work in Section 5.

## 2. Model Formulation

The DAD interdiction problem including defender's deception is formulated as a tri-level optimization problem in this section. In the first level, the defender seeks to minimize System Operating Cost ($SOC$) in addition to the required investment resources including hardening resources ($HR$), reinforcement resources ($RR$), and deception resources ($DR$) by allocating the limited resources to harden/reinforcement the grid's elements or propagating false information about hardening strategies. Note that hardening refers to a protection status that makes element protected against attack, and grid's elements are referred to the buses (substations), generators and transmission lines (power transformers also considered as the transmission lines) throughout this paper. The full knowledge of defender's decisions, which may include false information, is then passed to the second-level attacker problem who seeks to maximize $SOC$ by damaging the power network elements considering his limited budget. In the third level, the operator evaluates how the defender and attacker strategies influence the system operation. This process can be formulated by three hierarchical dependent optimization problems, where the formulation is discussed in the next sub-sections.

### 2.1  Deception Modelling

We provide the defender with an option to mislead the rivals about his plans. The importance of the deception strategy motivates the researchers in different fields of study to develop its formulation. Utilizing asymmetric information in the game-theoretic environment [24-29], using probabilistic approach [30, 31], graph theory [30], cosine similarity of binary variables [32], and incorrect cost function [29] are among the reported efforts to model the deception strategy. Considering the importance of feint strategy in the decision-making process, the works mentioned above try to model the deception and simulate the reaction of the opponent to the false data. In this paper, we introduce a novel formulation for deception and releasing disinformation based on the Boolean logic that enables the posturing strategy to be used in the resource allocation problem. Assume $\pi^D$ represents the real strategy made by the defender ($\pi^D = 1$ hardened, $\pi^D = 0$ not hardened element), and $\pi^{Df}$ is the strategies data disseminated to the attacker, where $\pi^{Df} = 1$ means the posturing strategy should be taken for the component. We define another variable $\pi^f$ to express the defender's effort for generating the false information. Fig.1(a) enumerates the possible defender's strategy for data propagation. For example, the first row stands for the situation that the defender generates false information ($\pi^f = 1$) on the status of the hardened facility ($\pi^D = 1$), and the attacker will be informed that the facility is not defended ($\pi^{Df} = 0$). In the second row, however, $\pi^f = 0$ and the real information will be passed to the attacker ($\pi^{Df} = \pi^D$). One can consider Fig.1(a) as an XOR operator depicted in Fig.1(b) which $\pi^D$ and $\pi^f$ are the inputs and $\pi^{Df}$ is the output. Accordingly, the algebraic expression of Fig.1 is presented in (1).

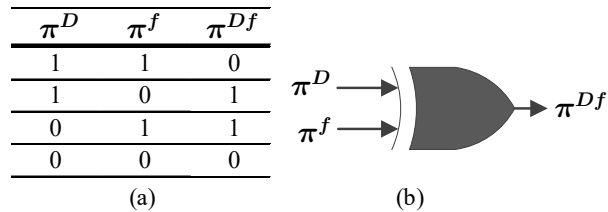

Figure 1.  (a) Binary representation of defender posturing. (b) XOR operator.

$$\pi^{Df} = (\pi^f + \pi^D) - 2\pi^f\pi^D \tag{1}$$

The above formulation paws the way for the defender to decide where the limited deceptive efforts should be allocated to have the maximum advantage in strategic planning.

*2.2     Shared Cognition Based Defender- Attacker- Operator Problem: Formulation*

The defender appropriates the available cyber-physical resources in two primary folds: physical resources and cyber resources. In physical resources, the defender distributes the hardening budget to secure the grid's elements based on the standard, e.g. NERC Physical Security CIP-014-2 [33]. In cyber resources, the defender employs the hardening for system protection such as protective hardware and software with an emphasis on prioritizing critical assets for protection, secure and harden configurations of power networks, and continuously assess for remediating vulnerabilities. Consequently, the defender needs to reinforce the physical and cyber resources to meet the required resilience level in the power grid. In the context of an entire economy, the resources mentioned above must be prioritized and allocated within the system to comply with the system security. Accordingly, the defender should emphasize to reduce *SOC* (2) and give the priority for allocation the *HR* (2) and *RR* (5) for the curtail network components, and *DR* (2) for the parts with lower importance. While the objective function (6) is a combination of different goals (2)-(5), the preemptive goal programming, discussed in subsection 3.2, will be employed to priorities the defender's goals.

$$SOC = \sum_{i \in B} C_i^{Sh} P_i^{Sh} + \sum_{g \in G} C_g P_g^G \tag{2}$$

$$HR = \sum_{i \in B} x_i^D + \sum_{(i,j) \in L} y_{ij}^D + \sum_{g \in G} z_g^D \tag{3}$$

$$RR = \sum_{(i,j) \in L} v_{ij}^D + \sum_{g \in G} w_g^D \tag{4}$$

$$DR = \sum_{i \in B} x_i^f + \sum_{(i,j) \in L} y_{ij}^f + \sum_{g \in G} z_g^f \tag{5}$$

The system operating cost (2) consists of the total power generation and total load shed cost. In general, the generation cost can be expressed by the quadratic function $\frac{1}{2} a_g P_g^2 + C_g P_g + b_g$. The marginal cost of generation unit $g$ could be calculated by $mc_g(P_g) = c_g(P_g) = a_g P_g + C_g$. In practice, $a_g$ is a very small number, therefore, the marginal cost could be presented by $mc_g(P_g) = C_g$ by ignoring the first term in the marginal cost function. $C_i^{sh}$ is the cost that shedding 1MW imposes on the system. Since load shed is not desired in our network, we consider a large value for it to act as a penalty in the objective function. Note that $C_i^{sh}$ must be pretty much larger than the largest marginal generation cost of a unit in the system. The Hardening resources (3) is the total utilized resources for hardening. The reinforcement resources (4) is the total spent reinforcement resources in the network, and the deception resources (5) is the total number of deceptive efforts that the defender adopted to secure his network. All these equations are positive, and the optimization problem tries to minimize each individual equation (2)-(5) to reach the lower value for the objective function (6). Note that all objective functions (2)– (5) are established based on the concept of resource allocation. In the context of operational research, a typical allocation problem involves the distribution of resources (e.g., generation dispatch, hardening, reinforcement, and deceptive resources in our model) among competing alternatives in order to minimize total costs. Such problems have the following components: a set of resources available in given amounts (e.g., minimum and maximum capacity of resources); a set of jobs to be done (e.g., enabling or disabling status of attacker/defender, load serving, and power conservation at each bus), each consuming a specified amount of resources; and a set of costs or returns for each job and resource.

Problem (6)-(22) has the tri-level structure, where the defender problem (7)-(12) is on the upper-level, the attacker problem (13)-(22) located at the middle-level, and the lower-level problem (17)-(22) is associated to the operator. Defender problem tries to minimize the cost function (6) with respect to the variables in set $\Delta^{\mathcal{D}} = \{x^D, y^D, z^D, v^D_{ij}, w^D_g, x^f, y^f, z^f, x^{Df}, y^{Df}, z^{Df}\}$. Equations (7)-(9) are modelling deception based on (1). Constraint (10) bounds the defender's resources for releasing false information. Similarly, (11) are the restrictions for the components' hardening, and constraints (12) are the limits on transmission lines and generators capacity increment. The attacker problem, which is in contrast to the defender problem, seeks to maximize cost function (6) with the decision variables in set $\Delta^{\mathcal{A}} = \{x^A, y^A, z^A\}$. Constraints (13)-(15) prohibit attacking to the defended elements. Note that in light of shared cognition approach, the defender can protect his assets by posturing rather than real protection. For instance, the defender may decide to use deception on the protection status of bus $i$, which is unprotected, i.e., $x^D_i = 0$. According to (7), when the defender sets $x^f_i = 1$ and uses a resource to make false information on the protection status of bus $i$, the transferred data to the attacker is $x^{Df}_i = 1$, that means the defender misleads the attacker on protection status of $x_i$. Then the deceived aggressor who received misinformation will not attack that bus due to the constraint (13). Constraints (16) limit the opponent's sources for striking the network component.

The last level operator problem which has the optimization variables in set $\Delta^{\mathcal{O}} = \{P^G, P^{Sh}, \theta, F\}$ evaluates both the attacker and defender tactics. Constraints (17) resembles the real part of the line power flow. Constraint (18) is the active power balance in each bus. Constraints (19) limit the permissible load-shedding to the amount of load. Constraint (20) is the power generator restriction. Constraint (21) limits the active power carried by the transmission lines. A transmission line is working with its full capacity, which is its fundamental capacity in addition to the capacity supplemented by the planner or is not functional because of damage to its own, or its two-ended buses as stated by equation (22). Moreover, with $\overline{F_{ij}}$ as the line capacity, we can provide the desired flexibility by considering the maximum line capacity, or merely a percentage of the full capacity by introducing a factor to go along with the capacity. The dual variables corresponding to the operator problem's constraint are presented after a colon.

$$\min_{\Delta^{\mathcal{D}}} \max_{\Delta^{\mathcal{A}}} \min_{\Delta^{\mathcal{O}}} (SOC + HR + DR + RR) \tag{6}$$

subject to

$$x^{Df}_i = \left(x^f_i + x^D_i\right) - 2x^f_i x^D_i; \quad \forall i \tag{7}$$

$$z^{Df}_g = \left(z^f_g + z^D_g\right) - 2z^f_g z^D_g; \quad \forall g \tag{8}$$

$$y^{Df}_{ij} = \left(y^f_{ij} + y^D_{ij}\right) - 2y^f_{ij} y^D_{ij}; \quad \forall (i,j) \tag{9}$$

$$\sum_{i \in B} x^f_i \leq R^{FB}; \sum_{(i,j) \in L} y^f_{ij} \leq R^{FL}; \sum_{g \in G} z^f_g \leq R^{FG} \tag{10}$$

$$\sum_{i \in B} x^D_i \leq R^{PB}; \sum_{(i,j) \in L} y^D_{ij} \leq R^{PL}; \sum_{g \in G} z^D_g \leq R^{PG} \tag{11}$$

$$\sum_{(i,j) \in L} v^D_{ij} \leq R^{RL}; \sum_{g \in G} w^D_g \leq R^{RG} \tag{12}$$

subject to

$$x^A_i \leq 1 - x^{Df}_i; \quad \forall i \tag{13}$$

$$z_g^A \leq 1 - z_g^{Df}; \quad \forall g \tag{14}$$

$$y_{ij}^A \leq 1 - y_{ij}^{Df}; \quad \forall (i,j) \tag{15}$$

$$\sum_{i \in B} x_i^A \leq R^{AB}; \sum_{(i,j) \in L} y_{ij}^A \leq R^{AL}; \sum_{g \in G} z_g^A \leq R^{AG} \tag{16}$$

subject to

$$F_{ij} = B_{ij}.(\theta_i - \theta_j).U_{ij}.\left(1 + \frac{c_{ij}v_{ij}^D}{\overline{F_{ij}}}\right); \forall(i,j) \; : \; (\lambda_{ij}^{\;F}) \tag{17}$$

$$\left(\sum_{g \in Gb_i} P_g^G - \sum_{(i,j) \in L} F_{ij} + \sum_{(i,j) \in L} F_{ji}\right) = P_i^D - P_i^{Sh}; \quad \forall i \; : \; (\lambda_i^{\;B}) \tag{18}$$

$$0 \leq P_i^{Sh} \leq P_i^D; \quad \forall i : \; (\mu_i^D) \tag{19}$$

$$0 \leq P_g^G \leq \left(\overline{P_g^G} + q_g w_g^D\right).(1 - z_g^A); \; \forall g \; : (\mu_g^G) \tag{20}$$

$$|F_{ij}| \leq (\overline{F_{ij}} + c_{ij}v_{ij}^D).U_{ij} \; ; \; \forall (i,j) : (\underline{\mu_{ij}^F}, \overline{\mu_{ij}^F}) \tag{21}$$

$$U_{ij} = \left(1 - x_i^A(1 - x_i^D)\right).\left(1 - x_j^A(1 - x_j^D)\right).\left(1 - y_{ij}^A(1 - y_{ij}^D)\right); \forall (i,j) \tag{22}$$

### 3. Solution Strategy

The multi-level format of the proposed approach in the foregoing section renders an NP-hard problem which is sophisticated to solve [9, 34], and the constraint-and-column generation (C&CG) method or a variant of Benders decomposition approach are usually employed to find the solution [35, 36]. In Bender's decomposition approach, the objective function of the so-called master problem is progressively constructed based on the dual information of the so-called sub-problem. Rather than using the dual information, the C&CG approach produces a set of primality cuts in each iteration that requires just primal variables. Since the continuous and differentiable sub-problem is not essential in the C&CG approach, and generally C&CG performs computationally better than its Benders' counterpart [36], the solution strategy is pursued by the C&CG method in this paper. The first step in solution strategy is decomposing the tri-level defender-attacker-operator problem into a master problem and a sub-problem. Since the operator is evaluating the defender and attacker strategies, we can form the master problem as the defender-operator (min-min), and the sub-problem as the attacker-operator (max-min) problems. The master problem has the min-min format that promises a single-level minimization problem, while the max-min sub-problem has the bi-level structure, that needs to be converted to the single-level one either through the KKT conditions or strong duality theorem. Note that our proposed formulas allow the transformation of the bi-level into the single level one because the operator problem is linear and thus convex on its decision variables. The overall procedure for the solution strategy, including the interactions among the sub-problems and master problem is schematically demonstrated in Figure 2.

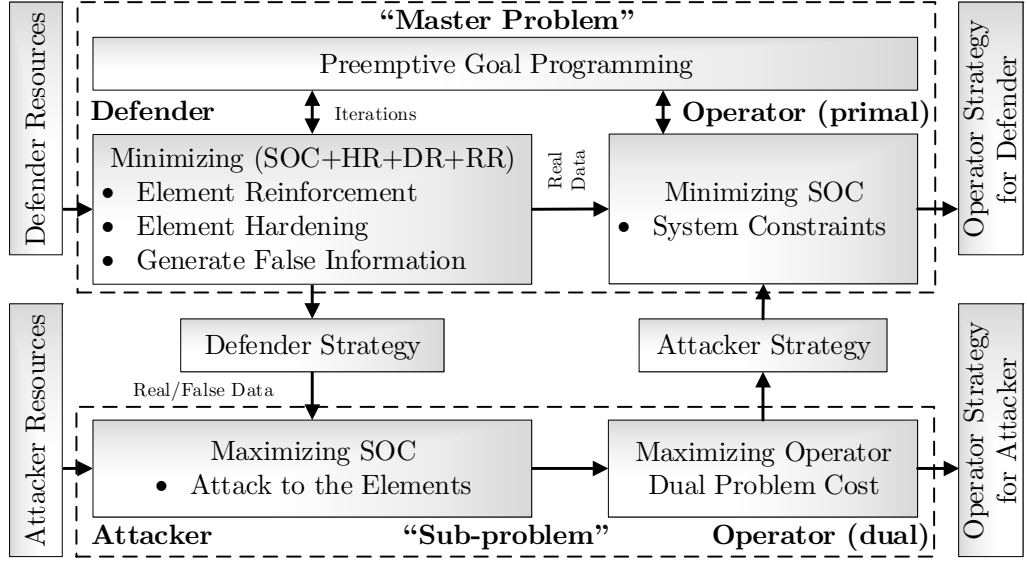

Figure 2. The procedure of the solution strategy

### 3.1 Master Problem

Master problem (24)-(36) is a minimization programming produced by the combination of defender and operator problems. The C&CG technique requires adding an iteration index $v$ (where $v = 1, \ldots, v^{max}$) to the optimization variables to facilate appending the new optimality cuts in each iteration.

$$\min_{\Delta^\mathcal{M}} \eta \tag{23}$$

subject to

$$\eta \geq SOC_v + HR_v + DR_v + RR_v \tag{24}$$

$$x_{i,v}^{Df} = \left(x_{i,v}^{f} + x_{i,v}^{D}\right) - 2x_{i,v}^{f}x_{i,v}^{D}; \quad \forall i \tag{25}$$

$$z_{g,v}^{Df} = \left(z_{g,v}^{f} + z_{g,v}^{D}\right) - 2z_{g,v}^{f}z_{g,v}^{D}; \quad \forall g \tag{26}$$

$$y_{ij,v}^{Df} = \left(y_{ij,v}^{f} + y_{ij,v}^{D}\right) - 2y_{ij,v}^{f}y_{ij,v}^{D}; \quad \forall (i,j) \tag{27}$$

$$\sum_{i \in B} x_{i,v}^{f} \leq R^{FB}; \sum_{(i,j) \in L} y_{ij,v}^{f} \leq R^{FL}; \sum_{g \in G} z_{g,v}^{f} \leq R^{FG} \tag{28}$$

$$\sum_{i \in B} x_{i,v}^{D} \leq R^{PB}; \sum_{(i,j) \in L} y_{ij,v}^{D} \leq R^{PL}; \sum_{g \in G} z_{g,v}^{D} \leq R^{PG} \tag{29}$$

$$\sum_{(i,j) \in L} v_{ij,v}^{D} \leq R^{RL}; \sum_{g \in G} w_{g,v}^{D} \leq R^{RG} \tag{30}$$

$$F_{ij,v} = B_{ij}(\theta_{i,v} - \theta_{j,v})U_{ij,v} \cdot \left(1 + \frac{c_{ij}v_{ij,v}^{D}}{\overline{F_{ij}}}\right); \forall (i,j) \tag{31}$$

$$\left(\sum_{g \in Gb_i} P_{g,v}^{G} - \sum_{(i,j) \in L} F_{ij,v} + \sum_{(i,j) \in L} F_{ji,v}\right) = P_{i,v}^{D} - P_{i,v}^{Sh}; \quad \forall i \tag{32}$$

$$0 \leq P_{i,v}^{Sh} \leq P_i^{D}; \quad \forall i \tag{33}$$

$$0 \leq P_{g,v}^{G} \leq \left(\overline{P_{g,v}^{G}} + q_g w_{g,v}^{D}\right)(1 - z_{g,v}^{A,*}); \quad \forall g \tag{34}$$

$$|F_{ij,v}| \leq (\overline{F_{ij,v}} + c_{ij}v_{ij,v}^{D}) \cdot U_{ij,v}; \quad \forall (i,j) \tag{35}$$

$$U_{ij,v} = \left(1 - x_{i,v}^{A,*}(1 - x_{i,v}^{D})\right) \cdot \left(1 - x_{j,v}^{A,*}(1 - x_{j,v}^{D})\right) \cdot \left(1 - y_{ij,v}^{A,*}(1 - y_{ij,v}^{D})\right); \forall(i,j) \qquad (36)$$

Set $\Delta^M$ contains the decision variables for the master problem including the auxiliary variable $\eta$ that gradually construct the cost function (23), the defender optimization variables $x_{i,v}^D, y_{ij,v}^D, z_{g,v}^D, v_{ij,v}^D, w_{g,v}^D$ and the optimization variables for the operator $P_{g,v}^G, P_{iv}^{Sh}, \theta_{i,v}, F_{ij,v}$. Each iteration of the master problem is supplied by a set of parameters $x_{i,v}^{A*}, y_{ij,v}^{A*}, z_{g,v}^{A*}$ that are the optimal values of the attacker decisions obtained from the sub-problem. Note that the " $*$ " superscript denoted the variable is fixed to its optimum value obtained from the other problem. Since the defender should make the prioritized decision to optimize the cost function (23), the preemptive goal programming is employed to distinguish between the posturing strategy and the other defensive attempts.

### 3.2 Preemptive Goal Programming Algorithm

The defender has the interest to minimize the *SOC* in addition to the investment cost for the equipment hardening/ reinforcement or false data propagation. Since an investor has a goal to protect the vital assets in the network with the hardening/reinforcement rather than posturing strategy, we employed the preemptive goal programming algorithm (PEPA) to make a difference among the real protection (hardening and reinforcement) and deception strategy. The basic purpose of goal programming is to simultaneously satisfy several goals relevant to the defender's decision-making problem. PEPA deals with the achievement of prescribed different goals or targets among diverse objective terms [37]. The order of priority is defined based on the high-cost impact of each goal. For example, given the order of preference in order as to minimize (1) the system operating cost, (2) hardening resources, (3) the reinforcement resources, and (4) deception resources. The PEPA is a well know algorithmic method to solve multi-objective optimization problems when each objective term has a contradiction with other terms; like Hardening and deception terms. One can multiply different weighting factors at each term of objective functions and solve the optimization problem. Obviously, this method does not give us a global optimal solution for the given weighting factors. Indeed, each selection provides an individual solution, which can represent as a Pareto front. The defenders can also do the same and set the weights and then solve the optimization problem and finally make a decision based on the Pareto front. Therefore, one can ask (like defender), what is the global optimum solution when we have no idea about the weights of the objective terms? In this section, we tried to answer this question. Thanks to the preemptive goal programming algorithm (PEPA) method to answer this question. This method follows a cumulative method to add each objective term one by one as it is elaborated in (37) – (47). Steps for the PEPA is provided as Algorithm 1.

---

**Algorithm 1** The PEPA algorithm. (For a given iteration $v$)

---

1: **Step 1**: Solve the master problem (23)-(36) with high prior objective function *SOC*, when

$$\min_{\Delta^M} \eta_1 \qquad (37)$$

subject to:

$$\eta_1 \geq SOC_v; \quad \forall v \qquad (38)$$

$$(25)\text{-}(36)$$

2: **Step 2**: If the first goal set is achieved, construct the new constraint (40) based on the achieved optimal solution of the problem in step 1, $\widehat{\eta_1}$, and add it to the following problem.

$$\min_{\Delta^M} \eta_2 \qquad (39)$$

subject to:

$$\widehat{\eta_1} + \epsilon_1 \geq SOC_v; \quad \forall v \tag{40}$$

$$\eta_2 \geq \epsilon_1 + HR_v; \quad \forall v \tag{41}$$

$$(25)\text{-}(36)$$

3: **Step 3**: Generate additional new constraint the same as steps 1 and 2 for the rest of the objective function terms sequentially as follows:

$$\min_{\Delta^{\mathcal{M}}} \eta_3 \tag{42}$$

subject to:

$$\widehat{\eta_2} + \epsilon_2 \geq HR_v; \quad \forall v \tag{43}$$

$$\eta_3 \geq \epsilon_1 + \epsilon_2 + RR_v; \quad \forall v \tag{44}$$

$$(25)\text{-}(36) \;\&\; (40)$$

Finally, for the last goal, the deception resources,

$$\min_{\Delta^{\mathcal{M}}} \eta_4 \tag{45}$$

subject to:

$$\widehat{\eta_3} + \epsilon_3 \geq DR_v; \quad \forall v \tag{46}$$

$$\eta_4 \geq \epsilon_1 + \epsilon_2 + \epsilon_3 + DR_v; \quad \forall v \tag{47}$$

$$(25)\text{-}(36) \;\&\; (40) \;\&\; (43)$$

4: **Step 4**: The final optimal value for the original objective function (23) in the PEPA is $\eta = \widehat{\eta_1} + \widehat{\eta_2} + \widehat{\eta_3} + \widehat{\eta_4}$

---

### 3.3 Sub-problem

Sub-problem is a combination of the attacker and operator problems that has the bi-level max-min structure. The linearity and convexity of the lower-level problem in its decision variables enables us to apply the strong duality theorem and obtain a dual operator problem that is a maximization problem. Achieving the max-max structure promises a single-level maximization structure for the Sub-problem, which formulated in equations (48)-(57). Note that the dual objective function (48) is equivalent to (6). However, since the attacker does not have any control on $HR, DR$, and $RR$, their optimal value entered as the parameter from the master problem.

$$\max_{\Delta^S} \left[ \sum_{g \in G} \mu_g^G \cdot \left(\overline{P_g^G} + q_g w_g^{D*}\right) \cdot (z_g^A - 1) - \sum_{(i,j) \in L} B_{ij} \cdot U_{ij} \cdot \left(1 + \frac{c_{ij} v_{ij}^{D*}}{\overline{F_{ij}}}\right) \cdot \left(\overline{\mu_{ij}^F} + \underline{\mu_{ij}^F}\right) + \sum_i P_i^D \cdot (\lambda_i^B - \mu_i^D) + HR^* + RR^* + DR^* \right] \tag{48}$$

subject to

$$x_i^A \leq 1 - x_i^{Df*}; \quad \forall i \tag{49}$$

$$z_g^A \leq 1 - z_g^{Df*}; \quad \forall g \tag{50}$$

$$y_{ij}^A \leq 1 - y_{ij}^{Df*}; \quad \forall (i,j) \tag{51}$$

$$\sum_{i \in B} x_i^A \leq R^{AB}; \sum_{(i,j) \in L} y_{ij}^A \leq R^{AL}; \sum_{g \in G} z_g^A \leq R^{AG} \tag{52}$$

$$C_g - \lambda_{i(g)}^B + \mu_g^G = 0; \quad \forall g \tag{53}$$

$$C_i^{Sh} - \lambda_{i(d)}^B + \mu_i^D \geq 0; \quad \forall i \tag{54}$$

$$\lambda_i^B - \lambda_j^B - \lambda_{ij}^F + \overline{\mu_{ij}^F} - \underline{\mu_{ij}^F} = 0; \quad \forall (i,j) \tag{55}$$

$$\sum_{j|(i,j) \in L} \lambda_{ij}^F B_{ij} U_{ij} \left(1 + \frac{c_{ij} v_{ij}^{D*}}{\overline{F_{ij}}}\right) - \sum_{j|(j,i) \in L} \lambda_{ij}^F B_{ji} U_{ji} \left(1 + \frac{c_{ji} v_{ji}^{D*}}{\overline{F_{ij}}}\right) \geq 0; \quad \forall i \tag{56}$$

$$\mu_i^D \geq 0, \forall i; \; \mu_g^G \geq 0, \forall g \; ; \; \overline{\mu_{ij}^F}, \underline{\mu_{ij}^F} \geq 0; \quad \forall (i,j) \tag{57}$$

where $\Delta^{\mathcal{S}} = \left\{x_i^A, z_g^A, y_{ij}^A, \lambda_i^B, \lambda_{ij}^F, \mu_g^G, \mu_i^D, \underline{\mu_{ij}^F}, \overline{\mu_{ij}^F}\right\}$. The dual cost function (48) is identical to the primal cost function (6). The attacker constraints (49)-(52) are equivalent to (13)-(16), while the defender's strategies are that parameters entered from the master problem. Constraints (53)-(57) are the stationarity equations in dual problem, and (57) entails some dual variable to be non-negative. Note that dual variable $\lambda_i^B$ in (53)-(55) adopts various subscripts $i(g), i(d), i, j$ which respectively stand for the bus that generator $g$ is located, the bus that load $d$ is located, the sending-end node of the ransmission line $(i, j)$, and the receiving-end node of transmission line $(i, j)$.

*3.4    Solution Algorithm for Bi-level Programming*

While the proposed sub-problem and the master problem has the non-linear formulation, they can be linearized according to the approaches explained in [34, 38]. Considering C&CG framework, lower-bound (LB) and upper-bound (UB) on the optimal value of objective function gradually constructs with the master problem and the sub-problem respectively. The optimal solution of the master problem at each iteration is entered as a parameter into sub-problem and vice versa. The iterative procedure lasts until the gap between the LB and UB is less than the predefined threshold $\delta$. The detailed steps of this iterative procedure are shown in Algorithm 2.

---

**Algorithm 2** The proposed bi-level programming algorithm

1: **Step 1** (Initialization): Set LB and UB bounds to $-\infty$ and $+\infty$, respectively. Set the iteration counter to $v = 0$. Set the attacker's variables $x_{i,v}^{A*}, y_{ij,v}^{A*}, z_{g,v}^{A*} = 0$.

2: **Step 2**: Update the iteration counter, $v \leftarrow v + 1$. Solve the master problem (23)-(36), using optimal values of attacker variables $x_{i,v-1}^{A*}$, $y_{ij,v-1}^{A*}$, $z_{g,v-1}^{A*}$ attained from Step 3 (or initialization) to be given parameter. In this step, the Algorithm 2 (PEPA algorithm) (37)-(47) should be called to find the original multi-objective function (23) at iteration $v$, in order to obtain optimal solution value of variables $\Delta^{\mathcal{M}*}$. Update LB as $\text{LB} = \eta^*$.

3: **Step 3**: Solve sub-problem (48)-(57) considering the optimal values of the defender's variables $x_{i,v}^{D*}, y_{ij,v}^{D*}, z_{g,v}^{D*}, v_{ij,v}^{D*}, w_{g,v}^{D*}$ to be given parameters. Obtain the optimal solution of sub problem's variables $\Delta^{\mathcal{S}*}$. Update UB using

$$\text{UB} = \min \left\{ \text{UB}, \left[ \sum_i P_i^D.(\lambda_i^{B*} - \mu_i^{D*}) + HR^* + \sum_{g \in G} \mu_g^{G*}.\left(\overline{P_g^G} + q_g w_g^{D*}\right).(z_g^{A*} - 1) + RR^* + DR^* - \sum_{(i,j) \in L} B_{ij}.U_{ij}^*.\left(1 + \frac{c_{ij}v_{ij}^{D*}}{\overline{F_{ij}}}\right).\left(\overline{\mu_{ij}^F}^* + \underline{\mu_{ij}^F}^*\right) \right] \right\}$$

4: **Step 4**: If $\text{UB} - \text{LB}$ is lower than a predefined tolerance $\delta$, terminate the algorithm and return the optimal solution in sets $\Delta^{\mathcal{S}*}$ and $\Delta^{\mathcal{M}*}$. Otherwise, continue in **Step 2**.

---

**4. Case Study**

The WSCC 9-bus and IEEE 118-bus systems are employed to demonstrate the performance of the proposed model. Note that the following considerations are made: $C_i^{Sh} = 1000$, $C_g$ is equal to the marginal cost of the quadratic generation cost function, the maximum iteration is $v^{max} = 50$ and the tolerance error is set $\delta = 10^{-5}$. The algorithm is implemented and executed using PC with an Intel® Core™ i5 CPU running at 3.2 GHz and 8GB of RAM using CPLEX under GAMS.

*4.1    WSCC 9-bus system*

The WSCC 9-bus system [39] with the load level of 315MW and optimum *SOC* equal to $28.4 is employed in this section to perform the various analysis. Since the model is designed to help the defender, in the first study, the attacker is assumed has a full resource for the attack on all elements. Then, the model performance is evaluated when the defender has a different number of hardening resources. To this end, three actions are considered which could be utilized by the defender; 1) Hardening, 2) Hardening and Reinforcement, 3) Hardening, Reinforcement, and Deception. The case studies for the 9-bus system are defined based on these actions as follows:

*Case 01: Hardening Action:* Hardening for 243 defender resource combinations; 243 different combination of defender resources (9 hardening resources for lines, 9 hardening resources for buses, 3 hardening resources for generators = 243) when the attacker has full resources (Figure 3).

*Case 02: Optimal Hardening Action:* Optimal hardening combination or optimum defense strategy (one combination out of 243 combinations is optimum). This optimum defender strategy is visualized in Figure 4(a). In this optimal hardening action when an attacker attacks any component, the system still has the normal operation.

*Case 03: Hardening & Reinforcement Action:* Figure 4(b) shows the optimum defense strategy with reinforcement action. It confirms that by increasing the capacity of generator G2 and its transformer at branch 2-8, the hardening is needed less equipment.

*Case 04: Hardening & Reinforcement & Deception Action:* In this case, we have three types of defending recourses. It turns out the problem to a multi-objective optimization problem, where two solution methodologies are possible; I) with PEPA, and II) without PEPA.

*Case 04-I) with PEPA:* In this method, we suppose that all resources are identical without any priority order to each other (Scenario 1 in Table 1).

*Case 04-II) with PEPA:* The main intention to use PEPA approach is to prioritize the hardening to deception resources (Scenario 2 in Table 1, and Table 2).

Figure 3 shows how the *SOC* changes with the different number of defender's resources for protecting line, buses, and generators in a 9-bus system. Figure 3 consists of three parts; in the upper part, surfaces associated with $R^{PG} = 1,2,3$ are coincident, while in the lower part only the associated surfaces with $R^{PG} = 2,3$ are coincident. The upper part of Figure 3 shows how the *SOC* changes by changing defender's resources for bus and line when only one generator G1, or G2, or G3 is protected. By increasing the defender's resources to $R^{PL} = 2$, $R^{PB} = 3$, and $R^{PG} = 1$ the model protects the highest load and cheapest generator by hardening generator {2}, buses {2, 8,9}, and lines {2-8,7-8}.

The middle surface shows hardening only generator G1 can promise lower *SOC* than protecting single generator G2 or G3. As can be seen, increasing defender's resources results in reducing *SOC*, however, when defender only protects one generator, cannot reduce *SOC* less than $45,033.1 ($33.1 generation cost plus $45,000 load shed cost), where buses {3,4,5,6,7,8,9}, and lines {3-6,4-5,4-9,5-6,6-7,7-8,8-9}, and generator {3} are hardened. The bottom surface shows the defender's strategies to omit load shed and promise minimum generation cost that can be achieved by protecting generators G2 and G3. The minimum element to be protected to achieve this goal is shown in Figure 4(a) where red components indicate the protected one. While reinforcement resources are not considered in the above-mentioned analysis, the effect of improving the capacity of lines and generator is shown in Figure 4(b), where defender adds 15MW to generator {2}, and 100MW to line {2-8}. By doing so, not only does the cost of operation reduces from $28.4 to $26.8, but fewer elements should be protected against attack. Both Figures 4(a) and 4(b) utilize the same formulation, namely (2)-(22), but the difference is the resources for capacity increment for the line ($R^{RL}$) and generator ($R^{RG}$) which are considered zero in the left system (Figure 4 (a)). The reason

why the PEPA approach is utilized in this paper is to prioritize the hardening to the deception. Since there are no deception resources in this case study, the solution can be obtained without using the PERA approach.

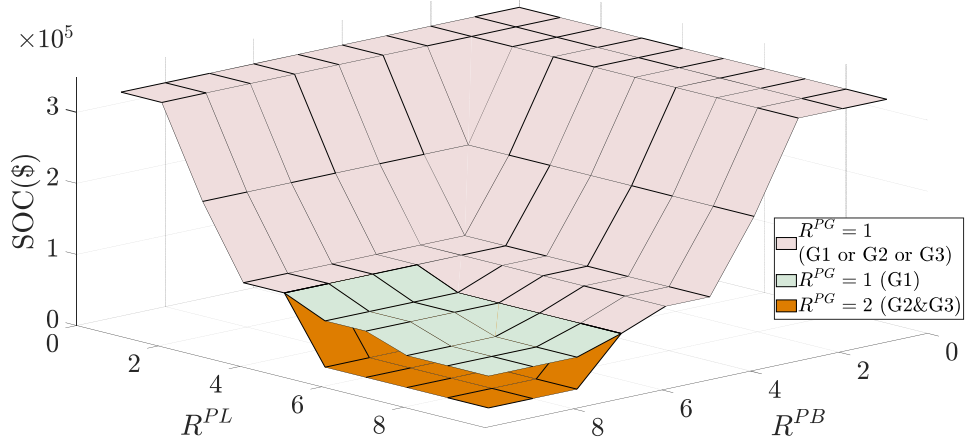

Figure 3. $SOC$ in different scenarios when $R^{PG} = 1,2,3$.

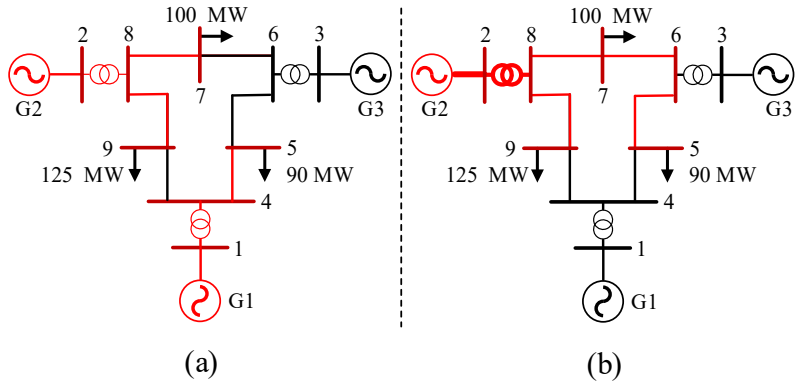

Figure 4. 9-bus system: (a) optimum defender's strategy without reinforcement. (b) Optimum defender's strategy with reinforcement of generator 2 and line 2-8.

Prioritizing resource allocation is essential for the defender who pursues the posturing strategy. Table 1 analyses the performance of the proposed model when the model solved without PEPA approach (scenario 1), and when the PEPA method is utilized (scenario 2). In both cases, the defender resources are $R^{PB} = 4$, $R^{PL} = 2$, $R^{PG} = 1$, $R^{FB} = 2$, $R^{FL} = 2$, $R^{FG} = 1$, $R^{RL} = 100$, $R^{RG} = 100$, while the aggressor has the full resource for the intrusion. In both scenarios, the defender decides to protect six buses, five lines, and one generator and the SOC will reach to the minimum possible value of $26.8. However, the defender does not have any preference to assign the deception to the part with the lower-importance in the first scenario, while in the second scenario, the defender allocates the hardening resources to the higher-priority equipment. Consequently, if the deception strategy leaked out and the invader attack to the not-hardened components, the damage imposed to the system in scenario 2 is much lower than scenario 1, as indicated in the column entitled "$SOC$ if Attack". The last column in Table 1," $SOC$ No Deception", represents the $SOC$ if no resources for deception strategy is available. Since both scenarios utilize the same number of resources for hardening, the $SOC$ is the same in both cases. It can be seen that $SOC$ increases if no deception applied, where the defender defends buses 2,8, 9, lines 8-2 and 8-9, and generator G2, and the attacker

attacks to other components. As can be seen, if no deception resources being available and the attacker intrudes the system, the *SOC* will increase by 52% and 117% for scenario one and two respectively.

**Table 1** Comparison between the solution with and without PEPA approach when attacker has full resources for attack.

| Scenario | $x_i^D$ | $z_g^D$ | $y_{ij}^D$ | $x_i^f$ | $z_g^f$ | $y_{ij}^f$ | $w_g^D$ | $v_{ij}^D$ | SOC if Attack | SOC No Deception |
|---|---|---|---|---|---|---|---|---|---|---|
| 1 | {2,5,7,8} | {2} | {4-5} | {4,9} | -- | {7-8,8-2, 8-9,9-4} | {2:15MW} | {8-2: 70MW} | $125,023 | $190,011 |
| 2 | {2,7,8,9} | {2} | {4-5, 8-2, 8-9, 9-4} | {4,5} | -- | {7-8} | {2:100MW} | {8-2: 100MW} | $90,022 | $190,011 |

**Table 2** Defender strategies when attacker has two resources for attacking to the buses, lines, and generators.

| $R^{PB}$ | $R^{PG}$ | $R^{PL}$ | $R^{FB}$ | $R^{FG}$ | $R^{FL}$ | $x_i^D$ | $z_g^D$ | $y_{ij}^D$ | $x_i^f$ | $z_g^f$ | $y_{ij}^f$ | $x_i^A$ | $z_g^A$ | $y_{ij}^A$ | LS (MW) | LSA (MW) |
|---|---|---|---|---|---|---|---|---|---|---|---|---|---|---|---|---|
| 0 | 0 | 0 | 0 | 0 | 0 | -- | -- | -- | -- | -- | -- | 1,3 | 1,3 | 6-7,8-2 | 315 | 315 |
| 1 | 1 | 1 | 0 | 0 | 0 | 2 | 2 | 8-9 | -- | -- | -- | 1,3 | 1,3 | 7-8,8-2 | 315 | 315 |
| 0 | 0 | 0 | 1 | 1 | 1 | -- | -- | -- | 2 | 2 | 8-9 | 1,3 | 1,3 | 7-8,8-2 | 315 | 315 |
| 1 | 1 | 1 | 1 | 1 | 1 | 2 | 2 | 8-9 | 9 | -- | 2-8 | 3 | 3 | 4-5,7-8 | 190 | 315 |
| 4 | 2 | 4 | 0 | 0 | 0 | 2,7,8,9 | 2 | 7-8,2-8,8-9 | -- | -- | -- | 1,3 | 1,3 | 4-5,6-7 | 90 | 90 |
| 2 | 1 | 2 | 2 | 1 | 2 | 2,9 | 2 | 2-8,8-9 | 8,7 | -- | 8-7 | 1,3 | 1,3 | 4-5,7-8 | 90 | 190 |
| 6 | 2 | 6 | 0 | 0 | 0 | 2,4,5,7,8,9 | 2 | 4-5,7-8,2-8, 8-9,4-9 | -- | -- | -- | 1,3 | 1,3 | 5-6,6-7 | 0 | 0 |
| 3 | 1 | 3 | 3 | 1 | 3 | 2,8,9 | 2 | 2-8,8-9,7-8 | 4,5,7 | -- | 4-5, 4-9 | 1,3 | 2,3 | 4-5,6-7 | 0 | 90 |

Table 2 demonstrates the defender's strategies given the various numbers of defensive resources, and when the attacker resources are $R^{AB} = 2, R^{AG} = 2, R^{AL} = 2$. The LS and LSA in the last two columns stand for the maximum load shed that respectively the deceived attacker and not-deceived attacker can impose to the network. The first row displays that if the network is defenseless, the attacker with only two resources for buses, lines, and generators can interrupt supplying the loads and cause 315 MW load shed. Even if the defender has one resource for protection grid's elements, he cannot prevent the full load shed within the system. Comparing the second and the third row conveys the fact that the defender protects the combination of bus 2, generator 2, and line 8-9 if he either has one set of hardening or deception resources for protecting the grid's elements. When the defender equipped with one set of resources for both hardening and deception, he can reduce the load shod to the 190 MW by deceiving the attacker. However, if the deception strategies leaked and the aggressor attacks to the bus 9 and line 2-8, he can cause the load shed of 315 MW.

The load shed is reduced to 90 MW when the defender has 4 resources for buses and lines, and two resources for generators. The defended elements are prioritized when the resources in the case mentioned previously are divided into hardening and deception resources. In this case, since the resources are enough for protecting the two loads, buses 2 and 9, lines 2-8 and 8-9, and generator 2 are hardened to ensure that the largest load (125MW) is protected, and buses 7 and 8 and line 7-8 are postured to be protected to save the 100MW load from the attack. Accordingly, in the case of deception failure, the damage would be minimized. There is no load shed in the system if the defender hardens 6 buses, 1 generator, and 5 lines, and the load shed slightly increases when the protecting resources are divided to hardening and deception in the last row, that implies the importance of using the PEPA approach in allocating the resources. This table also shows that generator 9, buses 2 and 9, and line 8-9 are among the top of hardening, while the buses 1 and 3 are mostly attacked. This counter-intuitive result helps the system planner to recognize the critical assets and make the judicious strategic planning for his network.

Our hardening and deception strategy are compared with the method proposed in [29], which utilized a Markov game model based on pre-defined five states with probabilities of defending and attacking to the lines. One of the five states is the "*static state*" without Markov model and probability, which our deterministic approach can be compared with. Since [29] only handles one attack and defend to the transmission lines at a time, we consider $R^{PL} = R^{AL} = 1$ to make the assessment. The comparison is shown in Figure 5 for three scenarios provided in [29] with static nature. The left part is identical to the result in [29], and the right part is calculated based on our proposed approach.

In Case I, line 2-4 is down and the approach proposed in [29] concludes that defender should defend line 2-8 and 1-4 with the probability of 81% and 19% respectively, and lines 2-8 and 5-6 are prone to attack with the chance of 32% and 38% respectively. In this case, in all four possible scenarios, there is no load shed in the system. However, our approach recognizes line 3-6 and 8-9 as the vulnerable component that has to be defended, otherwise damaging these lines cause 65MW load shed in the system. Since the outcome of the model should give an insight to the defender to recognize the components that are crucial for defense, our approach has privilege to the [29] due to recognizing lines 3-6 and 8-9 as a potential cause of 65MW load shed in the system, hence it brings the load saving of 65MW to the system.

In case II, lines 2-8 and 1-4 are recognized as the vulnerable component that their failure cause 65MW load shoed, however, our approach introduces lines 7-8 and 4-5 as the potential attacker target that can bring on 100 and 90 MW of load shed respectively. In this case, our approach promises at least 25MW load saving for the system should the attack succeed. While repairing line 1-4 in case III is an appropriate defending strategy adopted by [29], their method failed to recognize line 8-9 as a potential cause of 165 MW load shed in the system. While our deterministic approach is compared with the probabilistic method in [29], in overall, it can be seen that our proposed approach recognizes the more vulnerable lines that can provoke more amount of load shed in the system and adopts strategies that result in more load saving for the system in case of a successful intrusion. In most cases, the vulnerable components have not recognized with any chance as the vulnerable component in [29].

Our hardening approach for transmission lines has a similar result with [10]. However, both [10] and [29] only consider the transmission line as the target of attack and defend. As a more comprehensive model in comparison to [10], we further consider attacking the buses and generators, in addition to the fortification of transmission lines and generator units.

Figure 5. Comparison between the hardening methods proposed in [29] and our approach. The left part is based on [29], and the right part is based on our approach.

The deception strategy proposed in [29] assumes that the attacker will use a deceiving cost function to evaluate the cost of the load shed in the system, and their deception approach could gain the defender up to 50% of load shed reduction. In comparison the deception strategy proposed by [29], our approach has a more promising performance; we consider deception strategy not only for lines but for buses and generators; we exactly determine the component that should receive deception strategy based on the number of available deception resources. Table 3 demonstrates simple cases to show deception performance. In the first case, line 8-9 is down and the attacker with one resource attack to line 9-4 and cause 125MW load shed. Once the defender gets one deception resource, he defends line 9-4, and attacker damage declines by 48% to 65 MW. In Case 2, the deception resource can prevent any load shed in the system, gaining the defender 100% of load shed reduction.

**Table 3** Deception performance

| Case | Line Outage | $R^{PL}$ | $R^{FL}$ | $R^{AL}$ | $y_{ij}^D$ | $y_{ij}^f$ | $y_{ij}^A$ | LS(MW) |
|---|---|---|---|---|---|---|---|---|
| Case 1 | 8-9 | 0 | 0 | 1 | -- | -- | 9-4 | 125 |
|        | 8-9 | 0 | 1 | 1 | -- | 9-4 | 1-4 | 65 |
| Case 2 | 4-5 | 0 | 0 | 1 | -- | -- | 5-6 | 90 |
|        | 4-5 | 0 | 1 | 1 | -- | 5-6 | 2-8 | 0 |

*4.2    IEEE 118-bus system*

In this section, the proposed model is applied to the IEEE 118-bus system [39] with the load level of 4242 MW and optimum *SOC* of $59.1. Note that $\overline{F_{ij}}, \forall (i,j)$ are considered as 150 MW. The studies are performed to demonstrate the sensitivity of *SOC* to the defender's resources for protecting buses, generators and lines, considering the different amount of defender's resources for false information (posturing) in each case. Note that to obtain sensitivity analysis for one facility, for example, buses, the defender has ample resources for protecting the other facilities, e.g., lines and generators. Moreover, the attacker has full resources to attack all elements.

Figures 6 shows the sensitivity of the *SOC* to the number of protected buses ($R^{PB}$), number of protected lines ($R^{PL}$), and number of protected generators ($R^{PG}$). It also compares these sensitivities to the different number of false data (posturing) in each case. For example, Figures 6(a) shows how the *SOC* decreases when defender increases $R^{PB}$ (false data for protecting buses). Comparing blue and red lines show how the *SOC* decreases when the defender sends out 5 false data in protecting buses, and how *SOC* decreases further when $R^{FB}=10$, as represented by the green line. The difference between *SOC* for the blue line ($R^{FB}=0$) and the red line ($R^{FB}=5$) is represented by $\Delta Z^{0-5}$, which is the point-to-point difference in monetary gain that five more false data can bring to the system. Likewise, $\Delta Z^{5-10}$ is the point-to-point difference between the red and the green lines. It shows how much more saving can be achieved by increasing five more false data in addition to the hardening resources. The $\Delta Z^{0-5}$ and $\Delta Z^{5-10}$ can be viewed as the monetary worth of posturing or "value of posturing", i.e., the benefit that deception tactic brings to the system.

Figures 7 to 9 shows how adding more false data can bring more savings to the system. As can be seen, $\Delta Z^{0-5}$ is more effective than the $\Delta Z^{5-10}$ in most cases, meaning that "overdoing it" may not be beneficial in cost-saving. Moreover, the distribution of the $\Delta Z^{0-5}$ and $\Delta Z^{5-10}$ over $R^{PB}$, $R^{PL}$, and $R^{PG}$ gives an idea to the planner to how get the maximum benefits from posturing. For example, the maximum cost saving is achieved when defender utilized 5 more posturing resources while using 2 resources for hardening the buses ($\Delta Z^{0-5}$ at $R^{FB}=2$). The value of posturing for generators is depicted in Figure 7.

Surprisingly, the vast amount of posturing cannot reduce the objective function cost as a limited amount does. This figure shows a saving cost around to $2.5 million by applying only five deceptive resources, which is a considerable saving.

According to Figure 6 (a) and Figure 6 (b), while increasing the defender resources for hardening the facilities result in decreasing effect of the terrorist attack, the trend of this decrement is different in each facility. This fact speaks for the need for a comprehensive model, as we propose, to balance the deployment of different elements. More importantly, the value of posturing, as indicated in Figure 7 to Figure 9, implies different behavior considering the defender's resources for protection. For example, Figure 8 shows that the defender can get the highest benefit from posturing when he protects two buses and releases false information for 5 buses. As can be seen in Figure 6, there is a point that increasing number of defender resources for protecting buses, transmission lines, and generators does not lead to decreasing objective function more than $59.1 (optimal *SOC* in normal condition). These points are $R^{PB} = 97$, $R^{PL} = 89$, $R^{PG} = 38$. We have named this protection plan as the optimum defend strategy, where the system immune against any attack. This protection scheme is shown in Figure 11 (red elements are protected), which is obtained by running the model considering the above-mentioned defender's resources.

Figure 10 demonstrates the *SOC* and computational time in the IEEE 118-bus system, considering different values for $R^{PB}$ and $R^{PG}$, when the defender has full resources for protecting lines and the attacker has full resources for attacking all elements. Note that the tolerated relative optimality gap is set to 5%. Generally, it is shown that the problem is computationally tractable (the maximum running time is 1367 sec), and the running time is increased when $R^{PB}$ increase. Moreover, it is exposed that some combination of $R^{PB}$ and $R^{PG}$ need more computational burden.

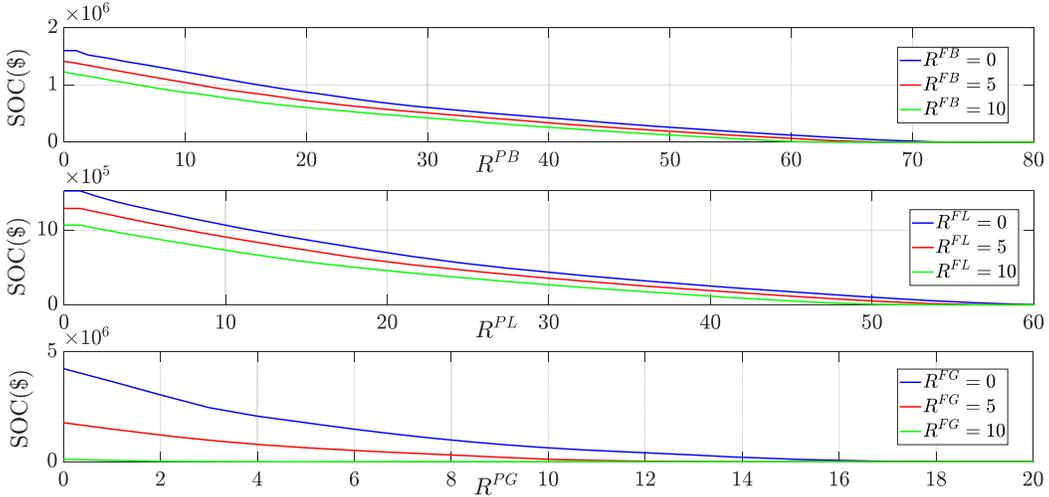

Figure 6. The sensitivity of ***SOC*** to defender resources for components hardening.

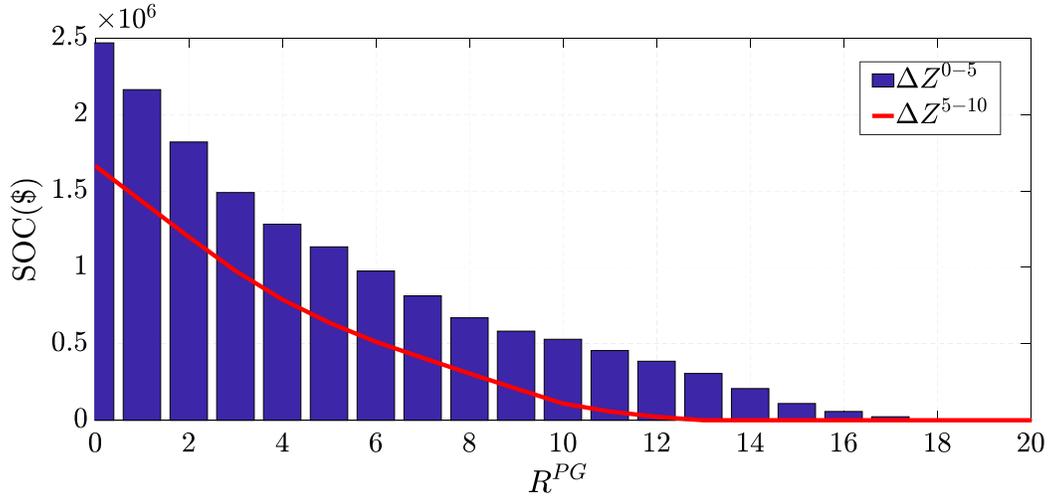

Figure 7. Value of posturing for protecting generators.

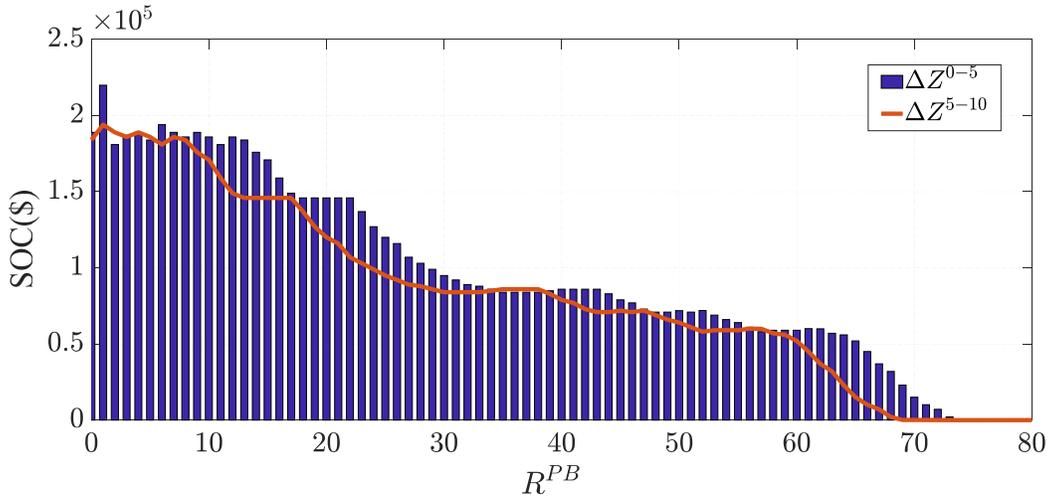

Figure 8. Value of posturing for protecting buses.

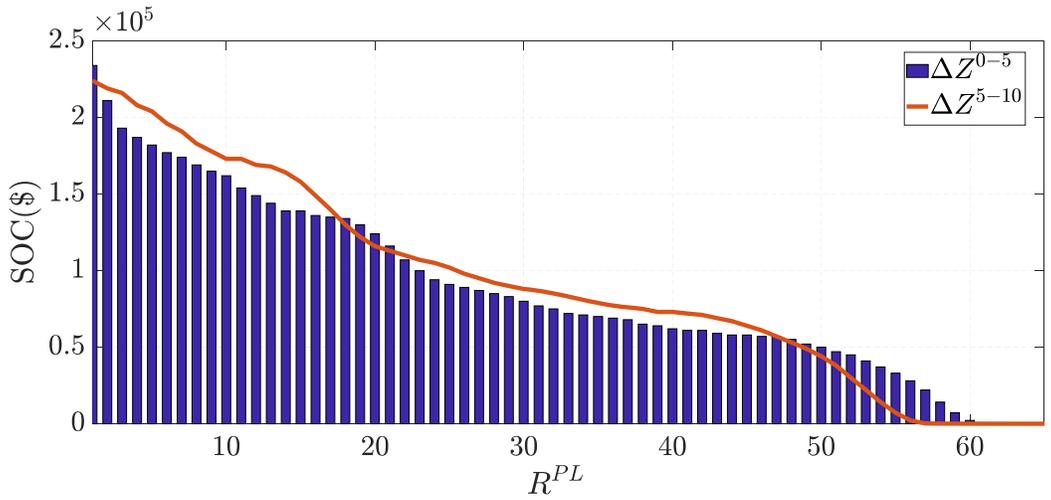

Figure 9. Value of posturing for protecting transmission lines.

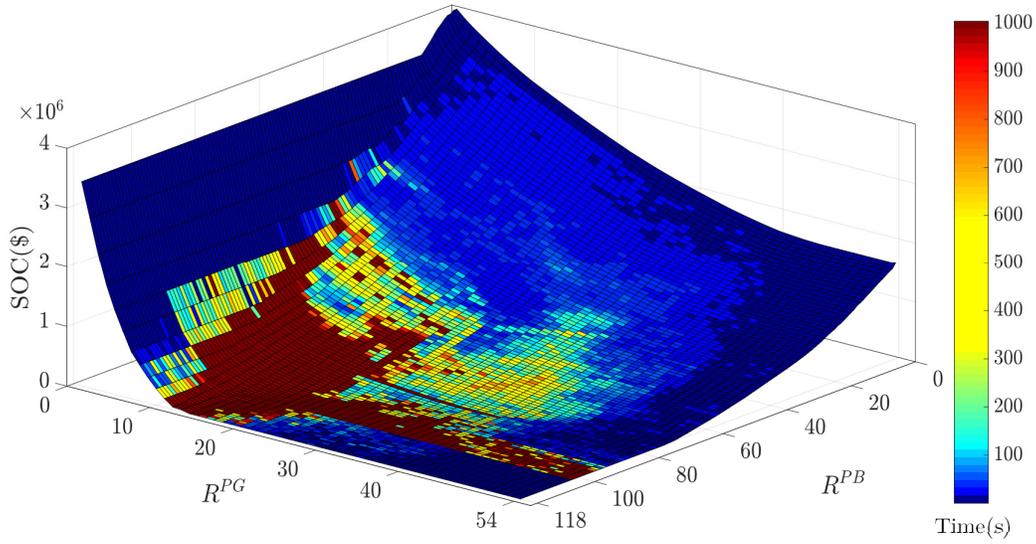

Figure 10. Computational time and $SOC$ with the different value of $R^{PG}$ and $R^{PB}$

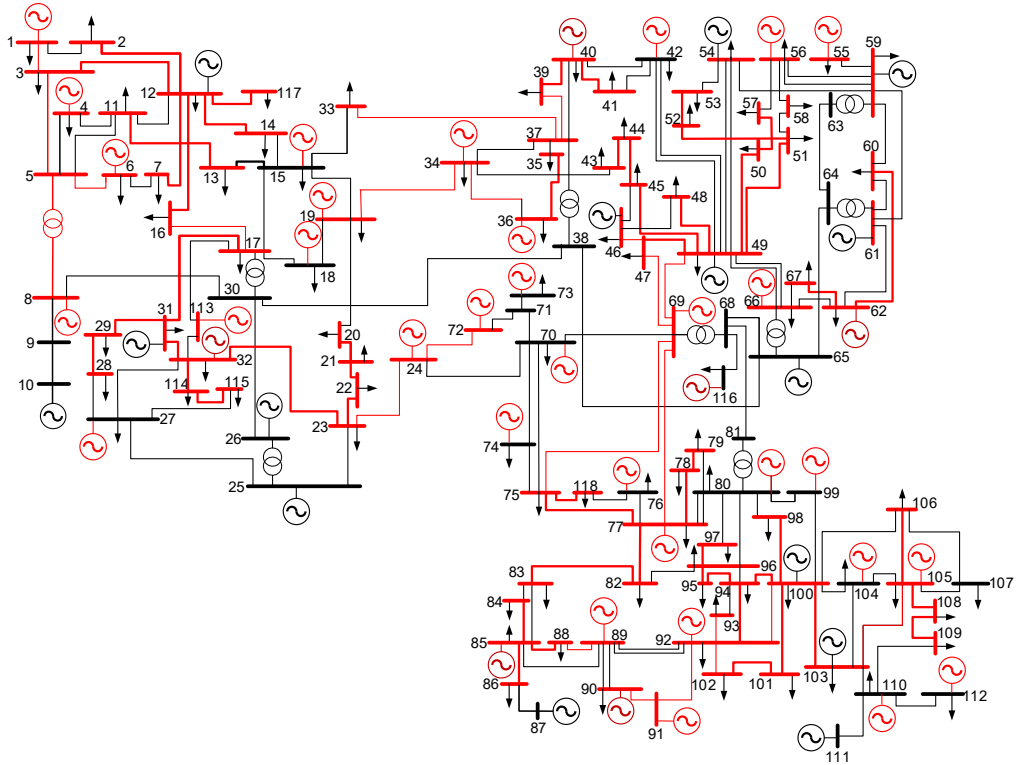

Figure 11. Optimum defend strategy in IEEE 118-bus test system. Red elements indicate protected one.

## 5. Conclusion

A multi-objective tri-level defender-attacker-operator problem to assist power system planner against deliberate attack is proposed in this paper. Attackers and defender can make protection and attack strategy on buses, transmission lines, and generators. Also, the defender has a choice to improve the transmission line and generator capacity as his defense strategy. Moreover, the defender can make a deception strategy rather than real protection (hardening) and deceive the attacker about the protection status of an element by posturing. A distinguishing feature of our model is our explicit recognition of the important information plays in such a competitive game, as shown in our discussions on shared cognition and epistemic knowledge, defined as knowledge supporting a belief, truth or hypothesis.

The tri-level problem is decomposed into a single-level defender-operator master problem and a bi-level attacker-operator sub-problem. Subsequently, the duality theorem utilized to transform the bi-level sub-problem to the single level one. The constraint-and-column generation technique is then employed to solve the problem, and the pre-emptive goal programming approach provides the defender with an option to protect the most critical facilities by the hardening strategy and applies the posturing tactic on the lower-priority parts. While the model is generally computationally tractable, the comprehensive case study shows the model has different behaviors in different cases, concerning computation burden and defender's strategy, highly dependent on the defender's resources for reinforcement, deception, and hardening. By performing various sensitivity analyses, also we reached the optimum defend strategy, which is a protection status of the power network that immunes the power system against any attack.

The monetary worth of deception is also calculated, that shows how much benefits deception strategy brings to the system with different protection status. It is shown that this amount is different in various strategies and the defender would be assisted by the model to make the best decision.

Accepted to be published in Journal of Electric Power Systems Research, 2018